\documentclass[aps,prb,twocolumn,superscriptaddress]{revtex4}
\usepackage{amsmath,amsfonts}
\usepackage{graphicx}
\PassOptionsToPackage{caption=false}{subfig}
\usepackage{subfig}
\usepackage[colorlinks=true,linkcolor=blue,citecolor=blue,linktoc=page]{hyperref}
\usepackage[all]{hypcap}
\usepackage{breqn}

\begin{document}
\title{Single and multi-particle scattering in Helical liquid with an impurity}
\author{Natalie Lezmy}
\affiliation{Department of Condensed Matter Physics$,$ Weizmann Institute of Science$,$ Rehovot$,$ 76100$,$ Israel}
\affiliation{Department of Particle Physics and Astrophysics$,$ Weizmann Institute of Science$,$ Rehovot$,$ 76100$,$ Israel}
\author{Yuval Oreg}
\affiliation{Department of Condensed Matter Physics$,$ Weizmann Institute of Science$,$ Rehovot$,$ 76100$,$ Israel}
\author{Micha Berkooz}
\affiliation{Department of Particle Physics and Astrophysics$,$ Weizmann Institute of Science$,$ Rehovot$,$ 76100$,$ Israel}
\begin{abstract}
We examine the scattering behavior from a single non magnetic impurity in a helical liquid. A helical liquid is a one dimensional system with a pair of counter propagating edge states, which are time reversal partners. In the absence of a magnetic field, time reversal symmetry is conserved and hence single particle scattering is prohibited. However, multi particle processes are possible. We examine the backscattering current and noise, and derive the effective scattering charge. A magnetic field enhances the backscattered current, but reduces the effective scattering charge. We find that the scattering charge can vary between $e$ and $2e$, depending on the strength of electron-electron interactions and the magnetic field.

\vspace{3mm}
\end{abstract}
\maketitle

\section{Introduction}

\begin{figure}[b]
\centering
\subfloat[single particle, $g_B$]{\label{fig:single}\includegraphics[scale=0.45]{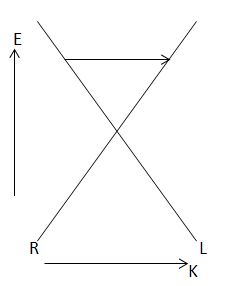}}
\subfloat[inelastic, $g_{ie}$]{\label{fig:inelastic}\includegraphics[scale=0.45]{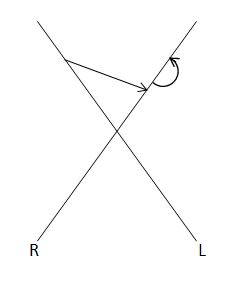}}
\subfloat[two particle, $g_{2p}$]{\label{fig:two}\includegraphics[scale=0.45]{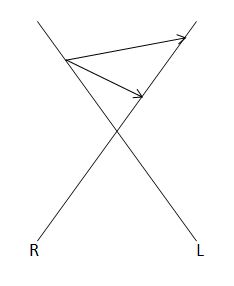}}
\caption{\label{fig:scattering} The scattering processes, on a diagram of the energy of the particles as a function of their momentum. In the single particle scattering, a left moving particle is scattered to a right moving particle with the same energy. The single particle process gives the strongest scattering, but allowed only in the presence of a magnetic field, which breaks time reversal symmetry. In the inelastic scattering , a left moving particle is scattered to a right moving particle of lower energy, and excites a particle-hole excitation on the right branch. In the two particle scattering, two left moving particles are scattered to two right moving particles, one with lower energy and the other with a higher energy. We see that both these scattering processes require more than one particle to conserve energy. These processes are allowed even without a magnetic field, but require electron-electron interaction. The strength of this interaction in the helical liquid determines which of the two is dominant.}
\end{figure}

A helical liquid is a topologically non trivial state of matter. It is a one dimensional (1D) system, with a pair of counter propagating edge states which are time reversal partners. Helical liquids have been realized on the edge of quantum spin hall insulators, in HgTe/(Hg,Cd)Te quantum wells\cite{kanemele2005,bernevig2006,konig2007}, and in InAs/GaSb Quantum Wells \cite{rui}. As long as time reversal symmetry is preserved, single particle backscattering (Fig. \ref{fig:single}), in which a left moving particle is scattered to a right moving particle (or the other way around), is prohibited. This endows topological stability to the edge states \cite{konig2008}. 

We will be interested in the dynamics of impurities embedded in the helical liquid. If we break time reversal symmetry at the impurity, which we will do using a weak magnetic field, single particle scattering will be enabled. The signature will be a backscattered current, which is the current reflected backward from the impurity, proportional to the magnetic field, with an effective scattering charge $e$. This process is characterized by the coefficient $g_B$.

However, even without a magnetic field, in the presence of electron-electron interaction, multi-particle scattering processes can occur and induce a backscattered current. These scatterings are inelastic, as more than one particle is necessary in order to conserve energy. Since these are multi particle processes, they will be less relevant at low energies than the single particle process. The emphasis of this paper are possible experimental signatures of the competition between weak multi particle processes and the stronger, but tunable, single particle process. In principle, the electron-electron interaction parameter is also tunable, although to a lesser extent.

In this work we will focus on the two lowest dimension multi particle scattering terms, each of which gives the dominant contribution to the backscattering at the appropriate range of the electron-electron interactions in the helical liquid. One is an inelastic scattering term, characterized by the coefficient $g_{\rm{ie}}$ (Fig. \ref{fig:inelastic}), in which, for example, a left moving particle is scattered into a right moving particle, and excites a particle hole excitation on the right branch. This operator represents several processes of this type, but in all of them a single unit of charge is backscattered and hence the scattering charge is $e$. The microscopic origin of this scattering term, in the weakly interaction case, was elucidated in a recent article by T. L. Schmidt \textit{et al.} \cite{schmidt}.

The other process involves two particle scattering \cite{bernevigprl2006}, characterized by the coefficient $g_{\rm{2p}}$ (Fig. \ref{fig:two}), where two left moving particles are scattered into two right moving particles (or the other way around). Since two particles are backscattered, the scattering charge is $2e$. 

The strength of these scattering processes depends on the IR energy scale, $E$, which is the bigger of the voltage and temperature. Looking at the derivative of the backscattering current with respect to the voltage bias, we find that
\begin{dmath}
\frac{dI}{dV}\simeq c_Bg_B^2E^{2K-2}+c_{ie}g_{ie}E^{2K+2}+c_{2p}g_{2p}^2E^{8K-2},
\label{eq:con}
\end{dmath}
where $K$ is the Luttinger parameter, which is a measure of the electron-electron interaction strength, and the $c$'s are numerical coefficients that would be calculated later. The non interacting case corresponds to $K=1$, and $K$ becomes smaller with increasing repulsive interaction, which is the case we focus on. The energy dependence is determined by the dimension of the operator - an operator of dimension $\Delta$ would contribute $E^{2\Delta}$ to Eq. (\ref{eq:con}), in first order perturbation theory. Relevant operators, in the renormalization group (RG) sense, localized at the impurity, have $\Delta<1$. In this case they become stronger, and induce a larger backscattered current at low energy. If $\Delta>1$, the power is larger than zero, and the backscattering is called irrelevant. We see that the two particle scattering becomes relevant at very strong interactions $K<\frac{1}{4}$, while the inelastic scattering is always irrelevant. We will work with electron-electron interactions in the range $\frac{1}{4}<K<1$, such that both the inelastic and two particle scattering are irrelevant.

In the absence of a magnetic field, the inelastic scattering term is dominant for weak electron-electron interactions $K>\frac{2}{3}$, \textit{i.e.} it has a lower dimension and hence it is stronger at low energy. For stronger interactions $K<\frac{2}{3}$ the two particle scattering is the dominant one. The single particle scattering term is always relevant, while the two other terms are irrelevant, and so it is always dominant at low energy scales. However, since its coefficient is proportional to the tunable external magnetic field, we can make it arbitrarily small, at a scale set by the voltage or temperature.

In this work, we consider the importance of each of the above scattering terms on the backscattering from a single, non magnetic, impurity. We apply a voltage bias on a helical liquid, to produce a current, and insert an impurity at $x=0$. We examine the backscattered current, noise and charge, as we change the energy scale, and how this behavior depends on the strength of electron-electron interactions, the magnetic field and the temperature. We find that at weak interaction, which is the case that has been realized in HgTe/(Hg,Cd)Te experiments, the scattering charge is $e$. For strong interaction, we find that the scattering charge can vary between $e$ and $2e$, depending on the rest of the parameters. These results are summarized in Fig. \ref{fig:eofvnob} and Fig. \ref{fig:eofvwb}. We also see that a magnetic field  greatly enhances the backscattered current.

The outline of the paper is the following. In section \ref{sec:model} we describe the model and the allowed operators for an arbitrary electron-electron interaction strength. We find the allowed operators, and show that they are indeed the lowest dimension ones, using the bososnic language to identify the quantum numbers scattering operators under time reversal (TR). In section \ref{sec:calc} (and in the supplemental material) we compute the current and shot noise, using the structure of the relevant conformal families \cite{bpz, difrancesco}, and the Keldysh formalism \cite{physicalkinetics,mahan,kamenev2009}. We work in the limit where the applied voltage bias is much larger than the temperature, and so the main contribution to the noise would be the shot noise and not the thermal noise. We do not include the constant thermal noise arising from the 1D mode, which exists in the absence of scattering. We then find the effective scattered charge from the relation between the current and the shot noise. Section \ref{sec:res} contains a representative sample of the behavior of the backscattered current, noise and charge. 

\section{Model}
\label{sec:model}

\subsection{Fermionic representation}
In this subsection we will present the model in terms of fermions, before proceeding to its bosonic justification in subsection \ref{sec:bos}.

The free Hamiltonian, in the absence of scattering terms, is \cite{wu2006}
\begin{dmath}
\mathcal{H}_{\rm{free}}=\int\!\mathrm{d}x\left[iv_f\psi_R^\dagger\partial_x\psi_R-iv_f\psi_L^\dagger\partial_x\psi_L+\frac{\pi}{2} \gamma\left(\rho_R+\rho_L\right)^2\right],
\end{dmath}
where $\psi^\dagger_{L(R)}$ is the single particle creation operator for a fermion moving left (right), and the density operator is  $\rho_{R\left(L\right)}\equiv\psi_{R\left(L\right)}^\dagger\psi_{R\left(L\right)}$.
The transformation of these fields under time reversal is \cite{konig2008}
\begin{dseries}
\begin{math}
\psi_L\rightarrow\psi_R
\end{math}
, \begin{math}
\psi_R\rightarrow-\psi_L.
\end{math}
\label{eq:tr}
\end{dseries}

After adding the impurity scattering, as we will argue in subsection \ref{sec:bos}, our Hamiltonian is
\begin{dmath}
\mathcal{H}=\mathcal{H}_{\rm{free}}+\mathcal{H}_{\rm{inelastic}}+\mathcal{H}_{\rm{2-particle}}+\mathcal{H}_{\rm{single}},
\label{eq:ham}
\end{dmath}
where
\begin{dgroup}
\begin{dmath}
\mathcal{H}_{\rm{inelastic}}=\tilde{g}_{\rm{ie}}\left(\partial\psi_L^\dagger\psi_L-\psi_R^\dagger\bar{\partial}\psi_R\right)\psi_L^\dagger\psi_R+h.c.\;\;
\label{eq:ferin}
\end{dmath}
\begin{dmath}
\mathcal{H}_{\rm{2-particle}}=\tilde{g}_{\rm{2p}}\psi_L^\dagger\partial\psi_L^\dagger\psi_R\bar{\partial}\psi_R+h.c.
\label{eq:fer2}
\end{dmath}
\begin{dmath}
\mathcal{H}_{\rm{single}}=\tilde{g}_{\rm{B}}(B)\psi_L^\dagger\psi_R+h.c.\;.
\label{eq:fer1}
\end{dmath}
\end{dgroup}
We emphasize that in the absence of a magnetic field, only the first two terms remain, \textit{i.e.} $\tilde{g}_{\rm{B}}(B=0)=0$. We use the derivatives $\partial=\partial_x+\frac{1}{u}\partial_t$ and $\bar{\partial}=\partial_x-\frac{1}{u}\partial_t$, where $u$ is the renormalized velocity defined in Eq. (\ref{eq:u}). These transform under TR as
\begin{dseries}
\begin{math}
\partial\rightarrow\bar{\partial}
\end{math}
, \begin{math}
\bar{\partial}\rightarrow\partial.
\end{math}
\end{dseries}
Using the TR transformation properties, we see that the above terms are time reversal even, provided that $\tilde{g}_B$ is an odd function of $B$.

It is convenient to work with dimensionless parameters, and so we will define
\begin{dgroup}
\begin{dmath}
\tilde{g}_{\rm{2p}}=g_{\rm{2p}}\frac{\hbar u}{a}(2\pi)^2a^4
\end{dmath}
\begin{dmath}
\tilde{g}_{\rm{ie}}=g_{\rm{ie}}\frac{\hbar u}{a}2\pi a^3
\end{dmath}
\begin{dmath}
\tilde{g}_{\rm{B}}=g_{\rm{B}}\frac{B}{B_0}\frac{\hbar u}{a}2\pi a.
\end{dmath}
\end{dgroup}
We work with a weak magnetic field $B$, and so in the last expression we focused on the leading, linear, term in it. $B_0$ is a normalization parameter, which we will take in our calculations to be $B_0=0.1T$. This is the strongest magnetic field we will look at, and we chose this $B_0$ following the experiment by Konig \textit{et al.} \cite{konig2008}.
 
In order to show that these are the most dominant scattering terms that satisfy the necessary symmetries, for arbitrary $K$, we will move to the bosonic representation of the system.

\subsection{Bosonic representation}
\label{sec:bos}
The effects of $\gamma$ can be handled in the Luttinger liquid approach, which will allow us to follow the operators (\ref{eq:ferin}-\ref{eq:fer1}) for any $\gamma$.
We will follow Giamarchi's bosonization scheme \cite{giamarchi}
\begin{dmath}
\psi_r\sim \frac{1}{\sqrt{2\pi a}}e^{-i(r\phi(x)-\theta(x))},
\end{dmath}
where $r$ is $+$ for $R$ and $-$ for $L$. The bosonized Hamiltonian is
\begin{dmath}
\mathcal{H}_{\rm{free}}=\int\mathrm{d}x\frac{u}{2\pi}\left[\frac{1}{K}\left(\partial_x\phi\right)^2+K\left(\partial_x\theta\right)^2\right],
\end{dmath}
where
\begin{dgroup}
\begin{dmath}
u=\sqrt{v_f^2+v_f\gamma}
\label{eq:u}
\end{dmath}
\begin{dmath}
K=\sqrt{\frac{v_f}{v_f+\gamma}}.
\end{dmath}
\end{dgroup}

After adding the impurity scattering, our Hamiltonian Eq. (\ref{eq:ham}) contains
\begin{dgroup}
\begin{dmath}
\mathcal{H}_{\rm{inelastic}}=\frac{\hbar u}{a}g_{\rm{ie}}a^2\left(\partial_x^2\theta\right)e^{2i\phi}(0)+h.c.
\label{eq:bosin}
\end{dmath}
\begin{dmath}
\mathcal{H}_{\rm{2-particle}}=\frac{\hbar u}{a}g_{\rm{2p}}e^{4i\phi(0)}+h.c.
\label{eq:bos2}
\end{dmath}
\begin{dmath}
\mathcal{H}_{\rm{single}}=\frac{\hbar u}{a}g_{\rm{B}}e^{2i\phi(0)}+h.c.\;.
\label{eq:bos1}
\end{dmath}
\label{eq:h}
\end{dgroup}

We will now show why these are the lowest dimension allowed scattering terms, up to total derivatives with respect to time. For this purpose, we will use conformal field theory formalism, which greatly simplifies the calculations.

In order to justify that these are the possible scattering terms and deal with them, it is more convenient to switch to the chiral description
\begin{dseries}
\begin{math}
\varphi_R=\theta-\phi
\end{math}
, \begin{math}
\varphi_L=\theta+\phi,
\end{math}
\end{dseries}
in which
\begin{dseries}
\begin{math}
\psi_L\propto e^{i\varphi_L}
\end{math}
, \begin{math}
\psi_R\propto e^{i\varphi_R}.
\end{math}
\label{eq:bosonization}
\end{dseries}
From a conformal field theory point of view, quantum numbers under TR are assigned to the primaries of the conformal representation. TR quantum numbers of the descendants are then fixed.
Since the TR transformation is discrete, it cannot change continuously, and so it remains the same as $K$ varies. Therefore, we can use the fermion representation to determine the TR transformation of the bosonic operators for arbitrary $K$.
From Eq. (\ref{eq:tr}), we have the TR transformation
\begin{dseries}
\begin{math}
e^{i\varphi_L}\rightarrow e^{i\varphi_R}
\end{math}
, \begin{math}
e^{i\varphi_R}\rightarrow-e^{i\varphi_L}.
\end{math}
\end{dseries}

More generally, consider the uncharged primary operators, which are $e^{-ip\varphi_L}e^{ip\varphi_R}$. We will only look at $p=1,2$, as higher values of $p$ correspond to terms of a higher dimension than we are interested in.
The TR transformation of the primary with p=1, which is the single particle scattering term is
\begin{dmath}
ce^{-i\varphi_L}e^{i\varphi_R}+h.c.=c\psi_L^\dagger\psi_R+c^*\psi_R^\dagger\psi_L\rightarrow-c\psi_L^\dagger\psi_R-c^*\psi_R^\dagger\psi_L=-c\left(e^{-i\varphi_L}e^{i\varphi_R}+h.c.\right),
\label{eq:p1}
\end{dmath}
where $c$ is some arbitrary coefficient. We see here that this term is odd under TR, and so it can only appear if we break time reversal symmetry.
The TR transformation of the primary with p=2, which is the two particle scattering term is
\begin{align}
ce^{-2i\varphi_L}e^{2i\varphi_R}&+h.c.\nonumber\\&=c\left(\partial\psi_L^\dagger\psi_L^\dagger-\psi_L^\dagger\partial\psi_L^\dagger\right)\left(\bar{\partial}\psi_R\psi_R
-\psi_R\bar{\partial}\psi_R\right)\nonumber\\
&+c^*\left(\bar{\partial}\psi_R^\dagger\psi_R^\dagger-\psi_R^\dagger\bar{\partial}\psi_R^\dagger\right)\left(\partial\psi_L\psi_L-\psi_L\partial\psi_L\right)\nonumber\\
&\rightarrow c^*\left(\bar{\partial}\psi_R^\dagger\psi_R^\dagger-\psi_R^\dagger\bar{\partial}\psi_R^\dagger\right)\left(\partial\psi_L\psi_L-\psi_L\partial\psi_L\right)\nonumber\\
&+c\left(\partial\psi_L^\dagger\psi_L^\dagger-\psi_L^\dagger\partial\psi_L^\dagger\right)\left(\bar{\partial}\psi_R\psi_R-\psi_R\bar{\partial}\psi_R\right)\nonumber\\
&=ce^{-2i\varphi_L}e^{2i\varphi_R}+h.c.\;.
\label{eq:p2}
\end{align}
We switch to the fermion representation here using Eq. (\ref{eq:bosonization}), with the standard substitution of $\psi_r\partial\psi_r$ instead of the composite operator of $\psi_r\psi_r$. We see here that this term is time reversal even and so it is a possible scattering term.

Even though the $p=1$ operator is odd under TR, some of its descendants would be TR even and may appear, even without a magnetic field. Our goal is to find the lowest dimension descendants that may appear. In order to get the descendants we act with lowering operators on the primaries. Lowering and raising operators are defined as \cite{difrancesco}
\begin{dmath}
L_n=\frac{1}{2\pi i}\oint dzz^{n+1}T(z),
\end{dmath}
and a similar definition for the right moving sector \footnote{Notice that we are working in a convention where $\bar{\partial}=\partial_x-\frac{1}{u}\partial_t$ in order to simplify signs under TR in the Virasoro algebra. In the more standard convention where $\bar{\partial}=\frac{1}{u}\partial_t-\partial_x$, the transformation laws would be $\bar{L}_n\rightarrow(-1)^nL_n$.}.
$T$ is the holomorphic part of the energy-momentum tensor, and in the right sector we have $\bar{T}$, the anti holomorphic part. By holomorphic (anti holomorphic), we mean the part that depends on $z=x+ut$ ($\bar{z}=x-ut$).

The lowering operators are $L_n$ with $n<0$. These operators transform under TR as $L_n\rightarrow\bar{L}_n$. Therefore, the level one (with dimension only raised by one) time reversal even term is
\begin{dmath}
c\left(L_{-1}-\bar{L}_{-1}\right)e^{-i\varphi_L}e^{i\varphi_R}+c^*\left(L_{-1}-\bar{L}_{-1}\right)e^{-i\varphi_R}e^{i\varphi_L}.
\end{dmath}
This term is even under TR since $L_{-1}-\bar{L}_{-1}\rightarrow\bar{L}_{-1}-L_{-1}$ and $e^{-i\varphi_L}e^{i\varphi_R}\rightarrow-e^{-i\varphi_R}e^{i\varphi_L}$.
However, this term is a total derivative with respect to time, and can therefore be neglected. When $L_{-1}$ $\left(\bar{L}_{-1}\right)$ acts on a primary \cite{bpz}, it acts as $\partial$ $\left(\bar{\partial}\right)$. Therefore, we can see that
\begin{dmath}[compact]
L_{-1}-\bar{L}_{-1}=\partial-\bar{\partial}=\frac{2}{u}\partial_t.
\end{dmath}

At level two we can have the following TR invariant operator
\begin{dmath}c\left(L_{-1}^2-\bar{L}_{-1}^2\right)e^{-i\varphi_L}e^{i\varphi_R}+c^*\left(L_{-1}^2-\bar{L}_{-1}^2\right)e^{-i\varphi_R}e^{i\varphi_L}\end{dmath}.
However, notice that
\begin{dmath}L_{-1}^2-\bar{L}_{-1}^2=\frac{4}{u}\partial_x\partial_t,\end{dmath}
and so this term is also a total derivative with respect to time, and can also be neglected.
The operator $L_{-1}\bar{L}_{-1}$ is even under TR, and so it would not give us a time reversal even descendant of the $p=1$ primary.
Finally, the last time reversal even term at level two is
\begin{equation}
\label{eq:l2}
c\left(L_{-2}-\bar{L}_{-2}\right)e^{-i\varphi_L}e^{i\varphi_R}+c^*\left(L_{-2}-\bar{L}_{-2}\right)e^{-i\varphi_R}e^{i\varphi_L}.
\end{equation}

In appendix \ref{app:deriv}, we explain that this term is not a total derivative with respect to time, and therefore it is the scattering term we wish to add. We also show that up to a total derivative with respect to time
\begin{dmath}c\left(L_{-2}-\bar{L}_{-2}\right)e^{-i\varphi_L}e^{i\varphi_R}+h.c.\propto c\left(\partial_x^2\varphi_L+\partial_x^2\varphi_R\right)e^{-i\varphi_L}e^{i\varphi_R}+h.c.\;.\end{dmath}
Or, moving back to the non-chiral form
\begin{dmath}c\left(L_{-2}-\bar{L}_{-2}\right)e^{-i\varphi_L}e^{i\varphi_R}+h.c.\;.\propto c\left(\partial_x^2\theta\right)e^{2i\phi}+h.c.\end{dmath}
We can also go back to the fermion representation
\begin{dmath}
c\left(\partial_x^2\varphi_L+\partial_x^2\varphi_R\right)e^{-i\varphi_L}e^{i\varphi_R}+h.c\propto c\left(\bar{\partial}\left(\psi_R^\dagger\psi_R\right)-\partial\left(\psi_L^\dagger\psi_L\right)\right)\psi_L^\dagger\psi_R+h.c.\;.
\end{dmath}
Thus, we have found all the possible scattering terms which are not of a higher dimension than 4 in the non interacting point, and we find the Hamiltonian in Eq. (\ref{eq:h}).

The time reversal transformation properties of the operators we examine are summerized in table \ref{tab:tr}.

\begin{table}[h]
\begin{tabular}{|p{3.5cm}|p{3.5cm}|}
\hline
Operator&After time reversal\\
\hline
$ce^{i\varphi_L}e^{-i\varphi_R}$&$-c^*e^{i\varphi_R}e^{-i\varphi_L}$\\
\hline
$ce^{2i\varphi_L}e^{-2i\varphi_R}$&$c^*e^{2i\varphi_R}e^{-2i\varphi_L}$\\
\hline
$cL_{-1}e^{i\varphi_L}e^{-i\varphi_R}$&$-c^*\bar{L}_{-1}e^{i\varphi_R}e^{-i\varphi_L}$\\
\hline
$cL_{-1}^2e^{i\varphi_L}e^{-i\varphi_R}$&$-c^*\bar{L}_{-1}^2e^{i\varphi_R}e^{-i\varphi_L}$\\
\hline
$cL_{-2}e^{i\varphi_L}e^{-i\varphi_R}$&$-c^*\bar{L}_{-2}e^{i\varphi_R}e^{-i\varphi_L}$\\
\hline
\end{tabular}
\caption{Transformation properties of operators under time reversal. The transformation of the primary operators is shown in Eq. (\ref{eq:p1}) and Eq. (\ref{eq:p2}). The transformation of the descendants is due to the transformation property $L_n\rightarrow\bar{L}_n$}
\label{tab:tr}
\end{table}

\begin{table*}[t]
\begin{tabular}{|p{1.3cm}|c|c|c|c|p{2cm}|}
\hline
&Fermionic&Chiral Bosonic&Non Chiral Bosonic&$\Delta$&Conductance voltage bias dependence\\
\hline
Inelastic&$g_{\rm{ie}}\left(\partial\psi_L^\dagger\psi_L-\bar{\partial}\psi_R^\dagger\psi_R\right)\psi_L^\dagger\psi_R+h.c.$&$g_{\rm{ie}}\left(\partial_x^2\varphi_L+\partial_x^2\varphi_R\right)e^{-i\varphi_L}e^{i\varphi_R}+h.c$&$g_{\rm{ie}}\left(\partial_x^2\theta\right)e^{2i\phi}+h.c.$&$K+2$&$V^{2K+2}$\\
\hline
Two \newline particle&$g_{\rm{2p}}\psi_L^\dagger\partial\psi_L^\dagger\psi_R\bar{\partial}\psi_R+h.c.$&$g_{2p}e^{-2i\varphi_L}e^{2i\varphi_R}+h.c.$&$g_{\rm{2p}}e^{4i\phi}+h.c.$&$4K$&$V^{8K-2}$\\
\hline
Single \newline particle&$g_{\rm{B}}\psi_L^\dagger\psi_R+h.c.$&$g_{\rm{B}}e^{-i\varphi_L}e^{i\varphi_R}+h.c.$&$g_{\rm{B}}e^{2i\phi}+h.c.$&$K$&$V^{2K-2}$\\
\hline
\end{tabular}
\caption{Scattering terms and their dimension. We find the dimension from the bosonic representation, using that the dimension of an operator of the form \cite{difrancesco} $e^{i\alpha\phi}$ is $\frac{\alpha^2}{4}K$, and a derivative adds 1 to the dimension. The voltage bias $V$ dependence of the conductance is $2\Delta-2$, where $\Delta$ is the dimension of the operator. This is for the case where the voltage bias is the relevant energy scale. We can easily see the dimension for $K=1$ from the fermion representation, as in that case the dimension of each fermion creation or destruction operator is $\frac{1}{2}$}
\label{tab:dim}
\end{table*}

In the bosonized form it is easy to deduce the dimension of the terms. Both forms of the scattering terms, their dimension, and the temperature dependence of the conductance they would give, are in table \ref{tab:dim}.

\begin{figure}[h]
\centering
\includegraphics[width=8.5cm]{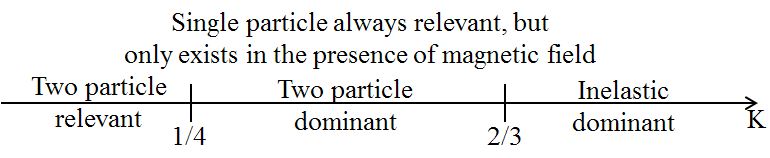}
\caption{For $K>\frac{2}{3}$, the dimension of the inelastic scattering term is lower than that of the two particle scattering term, and so it is dominant. When $K<\frac{2}{3}$, the dimension of the two particle term becomes lower, and so it gains dominance. However, as long as $K>\frac{1}{4}$, which is the parameter range we work in, both are irrelevant in the RG sense. At $K<\frac{1}{4}$, the two particle scattering term becomes relevant, while the inelastic scattering term always remains irrelevant}
\label{fig:dim}
\end{figure}

We see that the inelastic scattering term is irrelevant for all $K$, and the two particle scattering term is irrelevant for $K>\frac{1}{4}$. Therefore, for $K>\frac{1}{4}$, and in the absence of a magnetic field, both scattering terms are irrelevant. The single particle scattering term is relevant for $K<1$. We will examine systems where $\frac{1}{4}<K<1$. We also see that the inelastic scattering term is more dominant than the two particle scattering term for $K>\frac{2}{3}$. The Luttinger parameter ranges for the importance of the multi particle scattering terms is summarized in figure \ref{fig:dim}

Finally, we need to add the voltage bias to our Hamiltonian. We add the voltage bias by multiplying the scattering couplings by $e^{-\frac{ie^*Vt}{\hbar}}$, where $e^*$ is the effective charge of the scattering, and V is the voltage bias \cite{martin}. The complete Hamiltonian is then obtained from Eq. (\ref{eq:ham}) and Eq. (\ref{eq:h}) by the simple replacement $\phi\rightarrow\phi-\frac{ieVt}{2\hbar}$.

\section{Calculations}
\label{sec:calc}
The backscattering current in our system would be the change in the number of right moving particles multiplied by their charge
\begin{dmath}
I_b=e\frac{\mathrm{d}N_R}{\mathrm{d}t}=\frac{e}{i\hbar}\left[N_R,\mathcal{H}\right].
\end{dmath}
Therefore, we first need to write the particle number operator
\begin{dmath}
N_R=\int\mathrm{d}x\rho_R(x)=\int\mathrm{d}x\psi_R^\dagger(x)\psi_R(x)=-\frac{1}{2\pi}\int\mathrm{d}x\left(\partial_x\phi-\partial_x\theta\right).
\end{dmath}
Using this, we get the backscattering current operator
\begin{dmath}
I_b=\frac{ieu}{a}\left\{2g_{\rm{2p}}\left(e^{\frac{2ieVt}{\hbar}}e^{-i4\phi}-e^{\frac{-2ieVt}{\hbar}}e^{i4\phi}\right)
+g_{\rm{ie}}a^2\left(e^{i\frac{eVt}{\hbar}}\left(\partial_x^2\theta\right)e^{-2i\phi}-e^{-i\frac{eVt}{\hbar}}\left(\partial_x^2\theta\right)e^{2i\phi}\right)+g_{\rm{B}}\frac{B}{B_0}\left(e^{\frac{ieVt}{\hbar}}e^{-i2\phi}-e^{\frac{-ieVt}{\hbar}}e^{i2\phi}\right)\right\}.
\end{dmath}
Using the Keldysh formalism \cite{physicalkinetics,mahan,kamenev2009}, the average backscattered current is \cite{martin}
\begin{dmath}
\left\langle I_b(t)\right\rangle=\frac{1}{2}\sum_{\eta=\pm}\left\langle\hat{\mathrm{T}}_K\left[I_b(t^\eta)e^{-\frac{i}{\hbar}\int_K\mathrm{d}t_1\mathcal{H}(t_1)}\right]\right\rangle,
\end{dmath}
where $\hat{\mathrm{T}}_K$ is the Keldysh ordering operator, $t^{+(-)}$ is time $t$ on the upper (lower) branch of the Keldysh contour, and the integral is over the Keldysh contour.
The zeroth order term vanishes, and so we will work in first order perturbation theory.
We then need several correlation functions. There will be no term mixing the two particle scattering with the single and inelastic scatterings, as they have a different charge. In our calculation in the supplemental material we also show that a correlator mixing the inelastic scattering and single particle scattering term is zero as well. This makes sense, as the single particle scattering term has a linear dependence on the magnetic field, and so such a mixed term would also have a linear dependence on the magnetic field, which would contradict the Onsager relations, according to which the two terminal conductance is an even function of $B$. Therefore, we need three correlators, one for each of the three scattering terms with itself. We calculate these, using a conformal transformation to the plane, in the supplemental material, and eventually get \footnote{See supplemental material at ** for the detailed calculation of the correlation functions and the average backscattered current and noise}
\begin{dmath}
\left\langle I_b(t)\right\rangle\simeq I_{b_{\rm{2p}}}+I_{b_{\rm{ie}}}+I_{b_{\rm{B}}},
\end{dmath}
where
\begin{widetext}
\begin{subequations}
\begin{align}
I_{b_{\rm{2p}}}&=\frac{4eu}{a}g_{\rm{2p}}^2\sinh\left(eV\beta\right)\left(\frac{2\pi a}{u\beta\hbar}\right)^{8K-1}\frac{\left|\Gamma\left(4K+\frac{ieV\beta}{\pi}\right)\right|^2}{\Gamma(8K)},\allowdisplaybreaks\\
I_{b_{\rm{ie}}}&=\frac{2eu}{a}g_{\rm{ie}}^2\sinh\left(\frac{eV\beta}{2}\right)\left(\frac{2\pi a}{u\beta\hbar}\right)^{2K+3}\left[\frac{(1+22K)\left|K+1+\frac{ieV\beta}{2\pi}\right|^2}{(2K+3)(2K+2)}-\frac{9}{2}K\right]\frac{\left|\Gamma\left(K+1+\frac{ieV\beta}{2\pi}\right)\right|^2}{\Gamma(2K+2)},\allowdisplaybreaks\\
\rm{and}&\nonumber\\
I_{b_{\rm{B}}}&=\frac{2eu}{a}\left(g_{\rm{B}}\frac{B}{B_0}\right)^2\sinh\left(\frac{eV\beta}{2}\right)\left(\frac{2\pi a}{u\beta\hbar}\right)^{2K-1}\frac{\left|\Gamma\left(K+\frac{ieV\beta}{2\pi}\right)\right|^2}{\Gamma(2K)}.
\end{align}
\end{subequations}
\end{widetext}

The very low temperature limit of the current is
\begin{subequations}
\begin{align}
I_{b_{\rm{2p}}}&=\frac{4e}{\hbar}g_{\rm{2p}}^2\left(\frac{a}{u\hbar}\right)^{8K-2}\frac{1}{\Gamma(8K)}|2eV|^{8K-1},\\
I_{b_{\rm{ie}}}&=\frac{2e}{\hbar}g_{\rm{ie}}^2\left(\frac{a}{u\hbar}\right)^{2K+2}\frac{1+22K}{\Gamma(2K+4)}|eV|^{2K+3},\\
\rm{and}&\nonumber\\
I_{b_{\rm{B}}}&=\frac{2e}{\hbar}\left(g_{\rm{B}}\frac{B}{B_0}\right)^2\left(\frac{a}{u\hbar}\right)^{2K-2}\frac{1}{\Gamma(2K)}|eV|^{2K-1}.
\end{align}
\end{subequations}

The noise is also defined using the Keldysh formalism \cite{martin}
\begin{dmath}
S\left(t,t'\right)=\sum_\eta\left\langle\hat{\mathrm{T}}_K\left[I_b\left(t^\eta\right)I_b\left(t{'}^{-\eta}\right)e^{-i\int\limits_K\mathrm{d}t_1\mathcal{H}\left(t_1\right)}\right]\right\rangle-2\left\langle I_b\right\rangle^2.
\end{dmath}
We will be interested in the zero frequency noise
\begin{dmath}
S(\omega=0)=\int_{-\infty}^\infty\mathrm{d}tS(t,0).
\end{dmath}
This time, the zeroth order term does not vanish, and so we can work to that order. We calculate the zero frequency noise in a similar way to the average backscattered current, in the supplementary material, and find
\begin{dmath}
S(\omega=0)=2e\left\{I_{b_{\rm{2p}}}\coth\left(eV\beta\right)
+I_{b_{\rm{ie}}}\coth\left(\frac{eV\beta}{2}\right)+
I_{b_{\rm{B}}}\coth\left(\frac{eV\beta}{2}\right)\right\}.
\end{dmath}

From the backscattered current and noise we can finally find the effective scattered charge by
\begin{dmath}
e^*=\frac{S}{2eI}.
\end{dmath}

\section{Results}
\label{sec:res}

\begin{figure}[h]
\centering
\includegraphics[width=9cm]{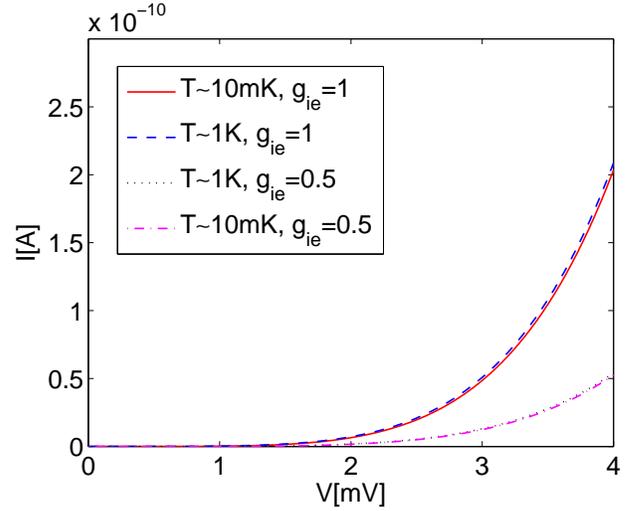}
\caption{Dependence of the backscattered current $I$ on the voltage bias $V$, without magnetic field. We vary $V$, and use the parameters $K=0.98$, $u=5.5\cdot10^5\frac{m}{s}$, $a=10^{-7}\mathrm{m}$, $g_{2p}=1$. We choose here to depict different values of $g_{\rm{ie}}$, rather than $g_{\rm{2p}}$, as for this Luttinger parameter it is the inelastic scattering which dominates the current.}
\label{fig:iofvnob}
\end{figure}

\begin{figure}[h]
\centering
\includegraphics[width=9cm]{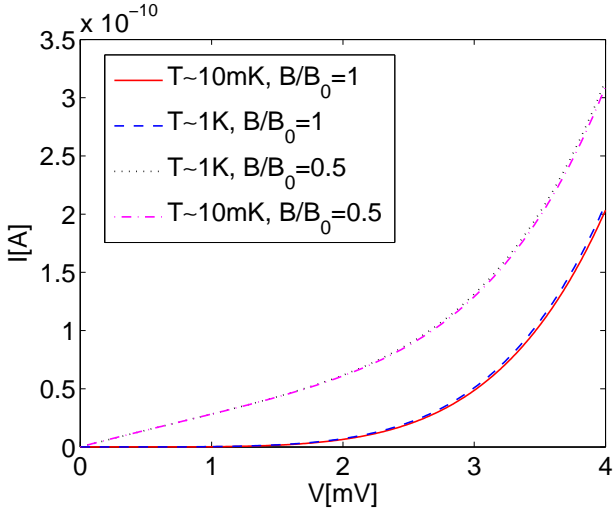}
\caption{Dependence of the backscattered current $I$ on the voltage bias $V$, in the presence of a magnetic field $B$. We vary $V$, and use the parameters $g_{\rm{B}}=0.02$, $B_0=0.1T$, $K=0.98$, $u=5.5\cdot10^5\frac{m}{s}$, $a=10^{-7}\mathrm{m}$, $g_{\rm{2p}}=1$, $g_{\rm{ie}}=1$. We see that the system is very sensitive to magnetic field, as the voltage dependence of the current is significantly modified. We also see that the main difference is at low voltage bias, where the single particle scattering term is more significant.}
\label{fig:iofvwb}
\end{figure}

In the experimental setup that we envision, there are 3 control parameters. Two of them are the voltage and temperature which influence the entire edge, and the third is a magnetic field which influences mainly the impurity properties. We work at an energy scale lower than the scale associated with the impurity, which we take to be $\frac{u\hbar}{a}$. This allows us to treat the impurity using the effective low energy theory.

Of the two time reversal even scattering terms, only one dominates the influence of the impurity at low energies, depending on the Luttinger parameter. At $K>\frac{2}{3}$, the inelastic scattering term, as in Eq. (\ref{eq:bosin}), is the dominant one, and at $\frac{1}{4}<K<\frac{2}{3}$ it is the two particle scattering term, as in Eq. (\ref{eq:bos2}). Both terms induce a backscattering current, each with a different scattering charge - $e$ for the inelastic scattering, and $2e$ for the two particle scattering. At zero magnetic field, the main result is the change in the effective scattering charge for different Luttinger parameters.

\begin{figure}[h]
\centering
\includegraphics[width=9cm]{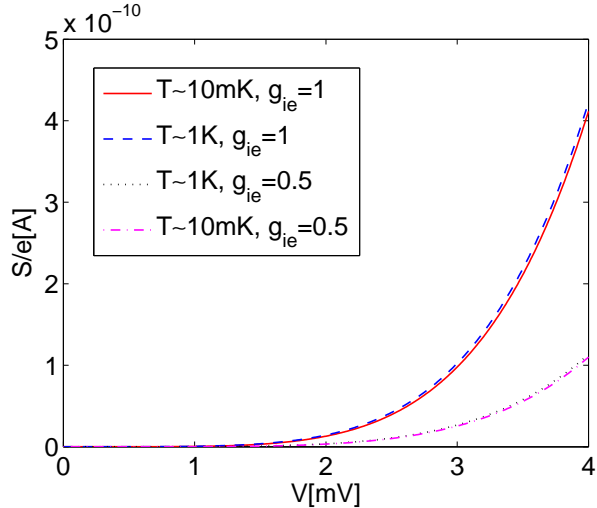}
\caption{Dependence of the noise $S$ on the voltage bias $V$, without magnetic field. We vary $V$, and use the parameters $K=0.98$, $u=5.5\cdot10^5\frac{m}{s}$, $a=10^{-7}\mathrm{m}$, $g_{\rm{2p}}=1$. Comparing to Fig. \ref{fig:iofvnob}, we see that the noise indeed seems to be proportional to the current}
\label{fig:sofvnob}
\end{figure}

\begin{figure}[h]
\centering
\includegraphics[width=9cm]{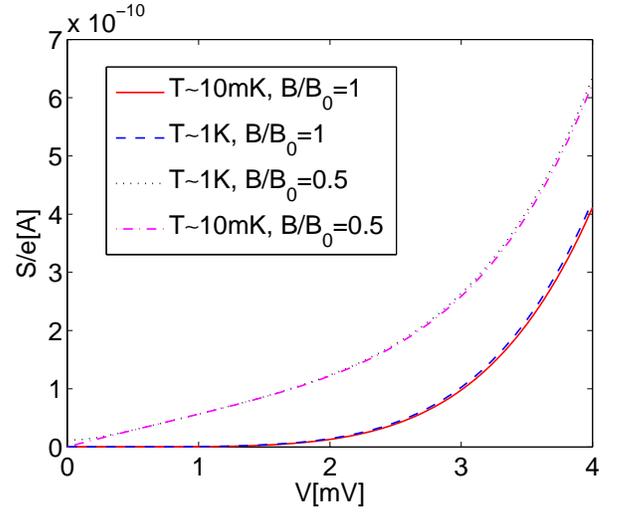}
\caption{Dependence of the noise $S$ on the voltage bias $V$, in the presence of a magnetic field $B$. We vary $V$, and use the parameters $g_{\rm{B}}=0.02$, $B_0=0.1T$, $K=0.98$, $u=5.5\cdot10^5\frac{m}{s}$, $a=10^{-7}\mathrm{m}$, $g_{\rm{2p}}=1$, $g_{\rm{ie}}=1$. Similarly to Fig. \ref{fig:iofvwb}, we see that the application of a magnetic field has a significant effect, especially at low voltage bias.}
\label{fig:sofvwb}
\end{figure}

\begin{figure}[h]
\centering
\includegraphics[width=9cm]{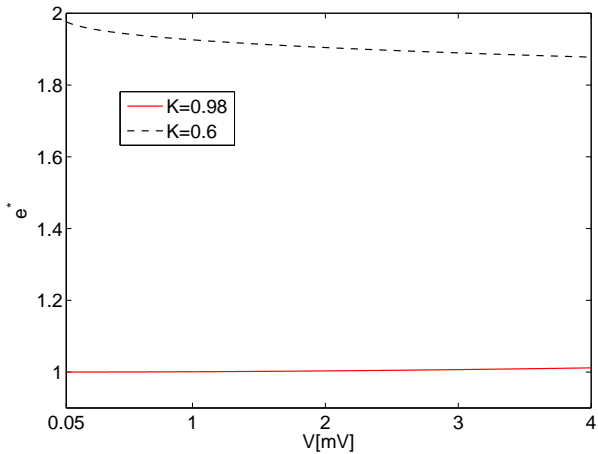}
\caption{Dependence of the effective scattered charge $e^*=\frac{S}{2eI}$ on the voltage bias $V$, without a magnetic field. We vary $V$, and use the parameters $u=5.5\cdot10^5\frac{m}{s}$, $T\simeq10\mathrm{mK}$, $a=10^{-7}\mathrm{m}$, $g_{\rm{2p}}=1$, $g_{\rm{ie}}=1$. We begin this graph at $V=0.05mV$, as we need to be in the parameter range where the voltage is much larger than the temperature, so that the main contribution to the noise is the shot noise, and not the thermal noise associated with the backscattering. We find that for weak electron-electron interactions, where $K$ is close to 1, the inelastic scattering term is dominant at low energies and so the scattering charge is $e$. At strong interactions, where $K<\frac{2}{3}$, the two particle scattering term dominates and so the scattering charge is close to $2e$. This is because the inelastic scattering term scales with the energy scale as $2K+2$, while the two particle scattering term scales as $8K-2$, see table \ref{tab:dim}.}
\label{fig:eofvnob}
\end{figure}

\begin{figure}[h]
\centering
\includegraphics[width=9cm]{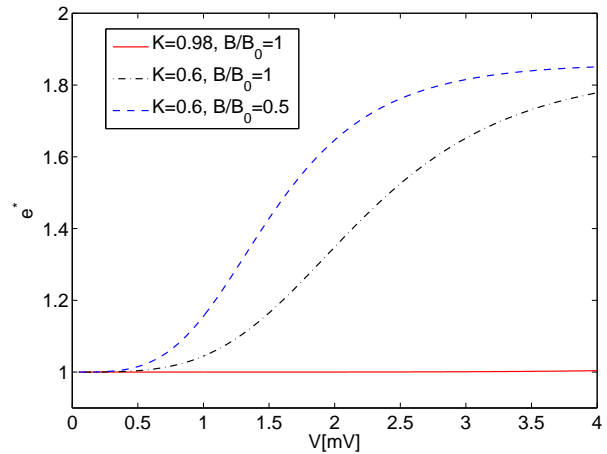}
\caption{Dependence of the effective scattered charge on the voltage bias, in the presence of a magnetic field. We vary $V$, and use the parameters $g_{\rm{B}}=0.02$, $u=5.5\cdot10^5\frac{m}{s}$, $T\simeq10\mathrm{mK}$, $a=10^{-7}\mathrm{m}$, $g_{\rm{2p}}=1$, $g_{\rm{ie}}=1$. We see that for weak electron-electron interactions $K=0.98$, where the inelastic scattering dominates over the two particle scattering, the effective scattering charge remains $e$. For this case, we do not depict different magnetic field strengths in this case, as there is no significant difference. For strong interactions $K=0.6$, where the two particle scattering dominates over the inelastic scattering, the charge is $e$ at low voltage bias, but increases toward $2e$ as the voltage is increased. This is because the two particle scattering term scales as $8K-2$ with the voltage, enhancing its effect at higher voltage, while the single particle scattering term scales as $2K-2$, decreasing its effect at higher voltage. We see that for a weaker magnetic field, the increase in the scattering charge is at a weaker voltage, and the ascent is more rapid}
\label{fig:eofvwb}
\end{figure}

When we turn on a magnetic field, the single particle scattering term, as in Eq. (\ref{eq:bos1}), is induced. This scattering term is a relevant operator for all $K<1$, and so  it is the dominant scattering term at low energies.  This scattering term gives a backscattered current, which is much stronger, at low energies, than the current due to the previous scattering terms. However, the scattering term is proportional to the magnetic field, and so we can control its strength. The scattering charge of this current is also $e$. As we change the voltage bias, the effective charge can vary between $e$ and $2e$, depending on the rest of the parameters, which are $K$, $T$, $g_{\rm{ie}}$, $g_{\rm{2p}}$, $g_{\rm{B}}$, $B$, $B_0$, $u$ and $a$.

This situation is depicted in the figures below. The figures contain two cases - the first is $K=0.98$ and the second is $K=0.6$. The first value of $K$ is motived by the experimental results of Konig \textit{et al.} \cite{konig2008}. This article also motivated the choice of the rest of the parameters - we chose their order of magnitude so the results roughly agree with the experimental results (this procedure should be taken with a grain of salt as it is possible that more than one impurity exists there). The case of $K=0.6$ has not been realized in an experiment so for, and so, for this case we maintained the rest of the parameters as those of the higher $K$ case. The parameters are therefore $g_{\rm{ie}}=1$ or $g_{\rm{ie}}=0.5$, $g_{\rm{2p}}=1$, $g_{\rm{B}}=0.02$, $B=0.1\mathrm{T}$ or $B=0.05\mathrm{T}$, $B_0=0.1\mathrm{T}$, $u=5.5\cdot10^5\frac{\mathrm{m}}{\mathrm{s}}$ and $a=10^{-7}\mathrm{m}$. Since the single particle scattering term is relevant, we must make sure the energy scale we work at is not so low that this scattering becomes too strong. When the energy scale is determined by the voltage, with $g_{\rm{B}}=0.02$ and $\frac{B}{B_0}=1$, we require $V>10^{-57}mV$ for $K=0.98$, which is a ridiculously small bound, and $V>0.04mV$ for $K=0.6$. If the temperature is the relevant energy scale, we need $T>10^{-44}\mathrm{K}$ for $K=0.98$, and $T>0.62\mathrm{K}$ for $K=0.6$. Since we also have irrelevant operators, we need to also make sure we do not go to an energy scale which is too high. For $g_{\rm{2p}}=g_{\rm{ie}}=1$, we need $V<10.5mV$ or $T<66\mathrm{K}$ for $K=0.98$, and $V<4.3mV$ or $T<54\mathrm{K}$ for $K=0.6$.

We start by looking at the dependence of the backscattered current on the voltage bias. Fig. \ref{fig:iofvnob} is for a system without a magnetic field, and Fig. \ref{fig:iofvwb} is for a system in the presence of a magnetic field. We see that even a weak magnetic field has a very significant effect on the backscattered current. A similar effect, of course, appears in the noise, depicted in Fig. \ref{fig:sofvnob} and Fig. \ref{fig:sofvwb}.

The effective scattered charge as a function of the voltage bias, for different Luttinger parameters, is depicted in figure \ref{fig:eofvnob} for a system without a magnetic field. We see that for $K>\frac{2}{3}$, the inelastic scattering term dominates at low energies, and so the effective scattering charge is close to $e$. For $\frac{1}{4}<K<\frac{2}{3}$, the two particle scattering term dominates, and pushes the effective scattering charge close to $2e$.

Figure \ref{fig:eofvwb} depicts the effective charge as a function of the voltage bias, for different Luttinger parameters and magnetic field strength. We see that for either value of $K$, the single particle scattering term is the dominant one at very low energies, and so at very weak voltage bias, the effective scattering charge is close to $e$. At $K>\frac{2}{3}$, the next to dominant term is the inelastic scattering term, and so the effective scattering charge remains close to e as we increase the voltage bias. For $\frac{1}{4}<K<\frac{2}{3}$, however, the two particle scattering term is the next to dominant term, and so increasing the voltage bias increases the effective scattering charge, taking it closer to $2e$.

\section{Summary}
In this work we examined the scattering from a single, non magnetic impurity, in a helical liquid. We discussed in detail the shot noise and the backscattered current due to the impurity. We discussed the importance of the inelastic, two particle and the single particle (which exists only after the application of a magnetic field) scattering processes in different regions of the parameters. The effective charge can approach $2e$ in certain situations, depending on the bias voltage $V$ and the Luttinger liquid interaction parameter $K$, see Fig. \ref{fig:eofvnob} and Fig. \ref{fig:eofvwb}.

We see that even though the single particle scattering term (discussed in Eq. (\ref{eq:bos1})) is the only relevant operator, the inelastic and two particle scattering terms are important, as the single particle scattering term vanishes at zero magnetic field and can be tuned by it. Of the two multi particle scattering terms, the inelastic scattering (discussed in Eq. (\ref{eq:bosin})) is dominant at weak electron-electron interaction $K>\frac{2}{3}$, and the two particle scattering (discussed in Eq. (\ref{eq:bos2})) is dominant at strong interaction $K<\frac{2}{3}$. We examine the effective scattering charge from the impurity, and see that in the absence of a magnetic field it is e for $K=0.98$, which is the case for HgTe/(Hg,Cd)Te quantum wells, and close to 2e for $K=0.6$. This is expected, as the inelastic scattering term has a scattering charge of $e$, and the two particle scattering term has a scattering charge of $2e$. In either case, at low energies, adding a magnetic field has a significant effect on the behavior of the backscattered current.

Applying a magnetic field gives a scattering charge of $e$ at low voltage bias for both interaction strengths, as it dominates at low energies. In the $K=0.6$ case, there is a range of voltage bias where the two particle scattering term is dominant, and so the effective scattering charge approaches $2e$. 

\begin{acknowledgments}
We thank A. Carmi for extremely helpful discussions, and C. Kane for emphasizing the importance of inelastic scattering. This research was supported by BSF and GIF grants, and an ISF center of excellence program.
\end{acknowledgments}

\appendix

\section{Descendant operators in derivative form}
\label{app:deriv}
The descendant operators of $e^{-i\varphi_L}e^{i\varphi_R}$ can also be expressed as polynomials of derivatives of $\varphi_L$ and $\varphi_R$ multiplying the primary. The TR transformation of the derivatives is
\begin{dmath}\partial^2\varphi_L=-\partial\left(\psi_L^\dagger\psi_L\right)\rightarrow-\bar{\partial}\left(\psi_R^\dagger\psi_R\right)=-\bar{\partial}^2\varphi_R\end{dmath}
\begin{dmath}\left(\partial\varphi_L\right)^2=-i\psi_L^\dagger\partial\psi_L+i\partial\psi_L^\dagger\psi_L\rightarrow i\psi_R^\dagger\bar{\partial}\psi_R-i\bar{\partial}\psi_R^\dagger\psi_R=\left(\bar{\partial}\varphi_R\right)^2,\end{dmath}
and so there are two time reversal even operators
\begin{dgroup}
\begin{dmath}
\left(\partial^2\varphi_L+\bar{\partial}^2\varphi_R\right)e^{-i\varphi_L}e^{\varphi_R}+h.c.
\end{dmath}
\begin{dmath}
\left[\left(\partial\varphi_L\right)^2-\left(\bar{\partial}\varphi_R\right)^2\right]e^{-i\varphi_L}e^{i\varphi_R}+h.c.\;.
\end{dmath}
\end{dgroup}

We can see that each of these terms on its own is not a total derivative with respect to time. However, there is a combination of them which is
\begin{dmath}
\partial_t\partial_xe^{-i\varphi_L}e^{i\varphi_R}=-i\partial_t\left(\partial_x\varphi_L-\partial_x\varphi_R\right)e^{-i\varphi_L}e^{i\varphi_R}= -i\partial_t\left(\partial_t\varphi_L+\partial_t\varphi_R\right)e^{-i\varphi_L}e^{i\varphi_R}=\left[-i\left(\partial_t^2\varphi_L+\partial_t^2\varphi_R\right)-\left(\left(\partial_t\varphi_L\right)^2-\left(\partial_t\varphi_R\right)^2\right)\right]e^{-i\varphi_L}e^{i\varphi_R},
\end{dmath}
and so these terms are equivalent up to a total derivative with respect to time.

Now we want to see how this connects with the lowering operator representation. In this dimension, we only have two such operators
\begin{dseries}
\begin{math}a\left(L_{-1}^2-\bar{L}_{-1}^2\right)e^{-i\varphi_L}e^{i\varphi_R}+h.c.\end{math}, and \begin{math}a\left(L_{-2}-\bar{L}_{-2}\right)e^{-i\varphi_L}e^{i\varphi_R}+h.c.\;.\end{math}
\end{dseries}
As we saw, the first is a total derivative with respect to time, and so the second must not be a total derivative with respect to time. Since these are the only operators, it must be equivalent to the term we found using the derivative representation.

\bibliographystyle{apsrev}
\bibliography{ti}
\end{document}